\documentclass[useAMS,usegraphicx,usenatbib]{mn2e}

\def\lesssim{\buildrel < \over {_{\sim}}}
\def\gtrsim{\buildrel > \over {_{\sim}}}

\title[Gamma ray emission from SNR RX J1713.7-3946 and the origin of
galactic cosmic rays]{Gamma ray emission from SNR RX J1713.7-3946 and the origin
of galactic cosmic rays}
\author[G. Morlino, E. Amato and P. Blasi]{
G. Morlino$^{1}$\thanks{E-mail: morlino@arcetri.astro.it}, 
E. Amato$^{1}$  \thanks{E-mail: amato@arcetri.astro.it}, and 
P. Blasi$^{1,2}$\thanks{E-mail: blasi@arcetri.astro.it; blasi@fnal.gov}, \\
$^{1}$INAF-Osservatorio Astrofisico di Arcetri, Largo E. Fermi, 5,
50125 Firenze, Italy\\
$^{2}$Center for Particle Astrophysics, Fermi National Accelerator
Laboratory, Batavia, IL, USA}

\begin{document}

\date{Accepted -----. Received -----}


\maketitle

\label{firstpage}

\begin{abstract}
We calculate the flux of non-thermal radiations from the supernova
remnant (SNR) RX J1713.7-3946 in the context of the non-linear theory of
particle acceleration at shocks, which allows us to take into account
self-consistently the dynamical reaction of the accelerated particles, the
generation of magnetic fields in the shock proximity and the dynamical reaction
of the magnetic field on the plasma. When the fraction of particles which get
accelerated is of order $\sim 10^{-4}$, we find that the strength of the
magnetic field obtained as a result of streaming instability induced by cosmic
rays is compatible with the interpretation of the X-ray emitting filaments
being produced by strong synchrotron losses in $\sim 100 \mu G$ magnetic fields.
The maximum energy of accelerated protons is $\gtrsim 10^5$ GeV. If the X-ray
filaments are explained in alternative ways, the constraint on the magnetic
field downstream of the shock disappears and the HESS data can be marginally fit
with ICS of relativistic electrons off a complex population of photons,
tailored to comprise CMB and ambient IR/Optical photons. The fit, typically
poor at the highest energies, requires a large density of target photons
within the remnant; only a fraction of order $\sim 10^{-6}$ of the
background particles gets accelerated; the local magnetic field is of order
$\sim 20\mu G$ and the maximum energy of protons is much lower than the
knee energy. Current HESS gamma ray observations combined with recent X-ray
observations by Suzaku do not allow as yet to draw a definitive conclusion
on whether RX J1713.7-3946 is an efficient cosmic ray accelerator, although
at the present time a hadronic interpretation of HESS data seems more
likely. We discuss the implications of our results for the GLAST gamma ray
telescope, which should be able to discriminate the two scenarios discussed
above, by observing the shape of the gamma ray spectrum at lower energies.
\end{abstract}


\section{Introduction}
 \label{sec:intro}

The detection of high energy gamma rays from supernova remnants has long been
considered as the most promising tool to test the supernova paradigm for the
origin of the bulk of cosmic rays observed at the Earth. Several supernova
remnants (SNRs) have now been detected in gamma rays, while useful upper
limits have been obtained for several others (see \cite{tati} and references
therein for a recent review of observations and their implications). 
The question now
to be answered is whether such emissions can be reliably considered as a
proof of the acceleration of hadronic particles (cosmic rays) or if they
should be rather interpreted as the result of radiative losses of relativistic
electrons. This goal can be achieved by investigating the following
aspects: 1) spectrum of the gamma ray emission combined with the spectrum
of X-rays (and possibly radio emission); 2) high energy cutoff of the gamma
ray emission; 3) morphology of the emission. 

Here we specialize our calculations to the SNR RX J1713.7-3946. 

For this remnant, the HESS Cherenkov gamma ray telescope has measured a
spectrum extending to $\sim 100$ TeV, although a power law spectrum with
an exponential cutoff at $E_\gamma^{max}$ leads to a best fit with
$E_\gamma^{max}\sim 10$ TeV \cite[]{aha06,aha07}. If this emission is due
to decay of neutral pions produced by nuclear collisions of relativistic
protons with some target gas in the shock vicinity, then this would imply 
a cutoff in the spectrum of accelerated protons at $\sim 100$ TeV, roughly 
a factor of 30 short of the knee. It is worth keeping in mind that in an 
expanding SNR the highest cosmic ray energy is expected to be reached 
around the beginning of the Sedov phase. Before
and after that time the maximum energy of accelerated particles does not
need to be as high as the knee energy. Moreover, different SNRs might
accelerate to different maximum energies because of the different
environment in which they explode. In this sense it is only natural that
most SNRs do not show proton acceleration up to the energy of the knee. 

The strength and topology of the magnetic field in the acceleration/radiation
region regulate the efficiency of particle acceleration (nuclei and electrons)
and the radiation losses of electrons. The recent detection of narrow X-ray
bright rims in several SNRs \cite[]{bamba03, bamba05, laz03, laz04, vink03}
has allowed to infer the strength of the field, if the thickness of the
rims is interpreted as the synchrotron loss length of  the radiating
electrons \cite[]{ballet}. Typical values of $50-1000\mu G$ have  been
inferred, which suggest efficient magnetic field amplification in the shock
region. One more argument in favor of strong magnetic fields was made
by \cite{uch07} who interpreted the rapid time variability (of the orders
of few years) observed in some X-ray emitting regions of RX J1713.7-3946 
as due to rapid synchrotron cooling. The field strength inferred based on 
this interpretation is as high as $\sim 1$ mG. The same effect was also observed 
in some knots of Cas A \cite[]{uch08}.

There is however a large source of ambiguity in that the narrow thickness of the
X-ray rims might be due to damping of the magnetic field (for instance due to
wave-wave non-linear coupling) \cite[]{pohl05}. In this case the magnetic field
inferred by assuming that the rims are due to severe synchrotron losses might
be overestimated. We discuss this issue at several points throughout the paper. 

Magnetic field amplification could result from turbulent amplification
\cite[]{joki} or from cosmic ray induced streaming instability. An open question
is whether the large magnetic fields present in the X-ray rims are widespread
throughout the remnant or are rather confined in thin filaments with a relatively
small filling factor (this would be the case if damping plays a role downstream
of the shock). This difference might have important consequences on the
interpretation of the observed non-thermal emission from SNRs, as we discuss
below. 

Interestingly enough, the magnetic field strength inferred from X-ray
observations is of the same order of magnitude as that required by the theory
of shock acceleration in order to account for particle acceleration up to the
knee region, provided the diffusion coefficient is Bohm-like. 

From the theoretical point of view, a major improvement has been achieved in 
the last few years in that the calculations are carried out in the
context of the so-called non-linear theory of particle acceleration. This
theory allows us to calculate the spectrum and spatial distribution of
accelerated particles around the shock, their dynamical reaction on the shock
and the magnetic field amplification that cosmic rays induce due to streaming
instability. 

Although the non-linear theory has been previously applied to SNRs, this paper
improves on previous attempts in two major aspects: 1) the magnetic field in the
shock region is calculated from streaming instability and from conservation
equations instead of being inserted {\it ad hoc} in such a way to fit the X-ray
observations; 2) we include the dynamical reaction of the turbulent magnetic
field on the shock, which is a very important effect in shaping the shock
precursor (\cite{cap08,long}).

The first point is of crucial importance in that it allows us to achieve a
unified picture of the spectra of X-ray and gamma ray emission, together with
the morphology of the X-ray emission, the strength of the magnetic field and 
finally the diffusion properties of the accelerated particles. Inclusion 
of the dynamical reaction of the field, then, as showed by \cite{cap08}, 
leads to a substantial reduction of the compression in the precursor, 
thereby also reducing the concavity of the spectrum of accelerated particles 
(\cite{long}). A reduced concavity of the spectrum should also be expected
as a result of the enhanced velocity of the scattering centers when the 
magnetic field is amplified (see for instance \cite{zir08}).

The large values of the magnetic field strength inferred from X-ray observations
and confirmed by our calculations clearly favor a hadronic interpretation of
the observed gamma ray emission, because a subdominant population of
electrons is sufficient to explain the spectrum and intensity of the observed
X-ray emission, but is insufficient to produce the observed gamma ray emission
through inverse Compton scattering (ICS) on the CMB and infrared/optical
photons (IR+Opt). The issue of whether the large fields are present in large
fractions of the remnant volume or are rather confined in narrow filaments,
remains however an open issue. 

For this reason we also investigate the possibility that the filaments have
some alternative explanation, so that there is no constraint on the strength of
the magnetic field downstream. We still use our non-linear calculations to
derive the spectra of accelerated protons and electrons, but force the
injection to be low enough to avoid large magnetic fields produced by streaming
instability. We find that a marginal fit to the gamma ray and X-ray data is
possible, although a large density of IR photons needs to be assumed inside 
the remnant. The fit to the highest energy part of the gamma ray
spectrum is not as good as for the hadronic interpretation.

The general conclusion we draw is that although there are numerous indications
that the observed gamma rays may be of hadronic origin, a clear answer on this
point may only come from: a) the extension of the gamma ray observations to
lower energies, and b) the detection of high energy neutrinos, which would be
produced in the same hadronic interactions \cite[]{neutrinos}.

The paper is organized as follows: in \S \ref{sec:model} we discuss in some
detail the general lines of the calculations we carry out, including the
non-linear theory of particle acceleration, the magnetic field amplification
and the different channels of non-thermal emissions. In \S \ref{sec:results} we
discuss the results we obtain by specializing the calculations to the case of
the SNR RX J1713.7-3946. We conclude in \S \ref{sec:conc}.

\section{Description of the calculations}
 \label{sec:model}

In this section we discuss the technical aspects of the calculations we carry
out, with special attention for the computation of the spectra of protons and
electrons, the magnetic field amplification and the radiation processes that
we include.

\subsection{Spectra of accelerated protons and electrons}
\label{sec:max_energy}

The dynamics of the shock region is dominated by accelerated protons, while
electrons are accelerated in the shock environment generated by protons. In this
sense the non-linearity of the problem is limited to the proton component,
which also generates the local magnetic field by exciting a streaming
instability. The spectrum of accelerated protons is computed using the 
method of \cite{amato05}. The acceleration time in
the presence of a precursor was determined by \cite{cap07}, and we
adopt that calculation. It was found there (and later confirmed for 
the parameters of RX J1713.7-3946 by \cite{vladi}) that the precursor 
reduces the maximum energy compared with the naive prediction of the 
test-particle theory with the same value of the magnetic field
upstream of the subshock.

The maximum momentum is first calculated using as a condition the equality
between the acceleration time and the age of the SNR. However we also
illustrate our results in the case in which the maximum momentum is
determined by the finite size of the accelerator. The diffusion coefficient
is chosen to be Bohm-like in the magnetic field generated through streaming
instability (see \S~\ref{sec:dyn}). Since the magnetic field depends on the
location in the precursor, the diffusion coefficient is also a function of
space: $D(p,x)=(1/3) p c^2/(e B(x))$. The two conditions mentioned above
read:
\begin{equation}
t_{acc}(p_{p,\rm max})=t_{SNR} \,\,\,,\,\,\,\,
\frac{D(p_{p,\rm max})}{u_0}=\alpha R_{SNR},
\label{eq:pmax}
\end{equation}
where $\alpha<1$ is the fraction of the shell radius ($R_{SNR}$) where the
escape of particles at $p\sim p_{p,\rm max}$ occurs. 

Although the basic structure of the calculation is the same proposed by
\cite{amato05} and \cite{amato06}, the crucial new aspect taken into account
here is the dynamical reaction of the self-generated magnetic field. We
introduce this effect following the treatment of \cite{cap08} and
\cite{long}: the conservation equations at the shock and in the precursor
are modified so as to include the magnetic contribution. The compression
factor at the subshock, $R_{sub}$, and the total compression factor,
$R_{tot}$, are deeply affected by this change, in that the ratio
$R_{tot}/R_{sub}$ (the compression factor in the precursor), decreases when
the amplified magnetic field contributes a pressure which is comparable
with the pressure of the thermal gas upstream. In turn this reflects in
spectra of accelerated particles which are closer to power laws, though a
concavity remains visible \cite[]{long}, as we discuss below.

The normalization of the proton spectrum is an output of our non linear
calculation, once a recipe for injection has been established. Following
\cite{bgv05}, particles are injected immediately downstream of the subshock.
The fraction of gas particles crossing the shock from downstream 
to upstream, $\eta_{\rm inj}$, can be written as
\begin{equation}
 \eta_{\rm inj}= 4/\left(3\sqrt{\pi}\right) 
 (R_{\rm sub}-1) \, \xi^3\,e^{-\xi^2}\,.
\label{eq:inj}
\end{equation} 
Here $\xi\sim 2-4$ is defined by the relation $p_{\rm inj}=\xi \, p_{\rm
th,2}$, where $p_{\rm th,2}$ is the momentum of the thermal particles
downstream. $\xi$ parametrizes the poorly known microphysics of the injection
process, but one should stress that $p_{\rm th,2}$ is an output of the problem:
as a result, the injection efficiency is affected by the dynamical reaction 
exerted by the accelerated particles and by the amplified magnetic field. 

The spectrum of accelerated electrons at the shock, $f_{e,0}(p)$, is easy
to calculate for $p\ll p_{e,\rm max}$, with $p_{e,\rm max}$ the
maximum electron momentum. In fact the slope of the electron spectrum at given
momentum $p$ is the same as the slope of the proton spectrum at the same
momentum, assuming that both electrons and protons experience the same
diffusion coefficient. The relative normalization of the two spectra
is unconstrained {\it a priori}, while the normalization of the proton
spectrum, as stressed above, is an output of our non linear calculation. 

The maximum energy of electrons is determined by equating the acceleration time
with the minimum between the time for energy losses and the age of the remnant.
The loss time of electrons over a cycle of shock crossing needs to be weighed by
the residence times upstream and downstream, therefore the condition for the
maximum momentum, in the loss dominated case, can be written as:
\begin{equation}
t_{acc}(p) = \left[ \tau_{l,1}(B_1,p)+\tau_{l,2}(B_2,p)\right]
\left[  
\frac{t_{r,1}(p)}{\tau_{l,1}(B_1,p)}+\frac{t_{r,2}(p)}{\tau_{l,2}(B_2,p)}
\right],
\label{eq:pemax}
\end{equation}
where $\tau_l$ denotes the loss time, and the indexes ``1'' and ``2'' refer to
quantities measured upstream and downstream respectively. The residence times
in the context of non linear theory of particle acceleration can be written
explicitely (from Eqs.~(25) and (26) of \cite{cap07}). Eq.~(\ref{eq:pemax})
cannot be solved for $p_{e,\rm max}$ analytically, contrary to the case of
acceleration in the test particle regime. The numerical solution is however
easy to obtain with standard techniques. 

When only synchroton losses are important, we can provide an approximate
expression for the solution of Eq.~(\ref{eq:pemax}). We use the function
$u_p(p)$ defined as the mean plasma velocity that a particle with momentum
$p$ experiences in the precursor region (see Eq.~(8) in \cite{amato05}).
The function $U_p(p)\equiv u_p(p)/u_0$ is such that $U_p(p)< 1$ for any
momentum. The electron maximum momentum can then be written in an implicit
way as:
\begin{equation} \label{eq:pmax_e_2}
 p_{e,\rm max}= \frac{3}{2} \sqrt{\frac{m_e^3 c^4}{e B_1 r_0}}\,
 \frac{u_0}{c}\, U_p(p_{e,\rm max})\, \sqrt{\frac{1-R_{\rm tot}^{-1}\,
 U_p^{-1}(p_{e,\rm max})}{1+R_B R_{\rm tot} U_p(p_{e,\rm max})}} \,,
\end{equation}
where $r_0$ is the classical electron radius. One may notice that 
Eq.~(\ref{eq:pmax_e_2}) reduces to the Eq.~(11) of \cite{ber06} if one
poses $U_p=1$. Eq.~(\ref{eq:pmax_e_2}) leads to an error of less than $10\%$
with respect to Eq.~(\ref{eq:pemax}) (but we stress that the results
presented in this work are all obtained using Eq.~(\ref{eq:pemax})).

The spectrum of electrons at energies around and above $p_{e,\rm max}$,
namely the shape of the cutoff is harder to calculate in the context of a
fully non linear calculation. In the test particle case and for strong
shocks the solution has been calculated by \cite{zir07}. Since the spectra
we find for electrons at $p<p_{e,\rm max}$ are not far from being
power laws with slope $\sim 4$, we adopt the modification factor calculated
by \cite{zir07} and we write the electron spectrum at the shock in the loss
dominated case as: 
\begin{equation} \label{eq:f_e(p)}
 f_{e,0}(p)= K_{ep}\,f_{p,0}(p)
  {\left[1+0.523 \left(p/p_{e,\rm max}\right)^{\frac{9}{4}}\right]}^2
  e^{-p^2/p_{e,\rm max}^2} \,.
\end{equation} 
The most important aspect of this expression is the fact that the cutoff is not
a simple exponential, which in turn reflects in the shape of the secondary
emission radiated by electrons. This important conclusion differs from
the numerous calculations previously carried out by other authors, where an
exponential cutoff was assumed for the electron spectrum. On the 
other hand, if the maximum momentum is indeed determined by the age of the
remnant, then the cutoff shape is expected to be exponential. The constant 
$K_{ep}$ accounts for the different normalization between the electron and 
proton spectrum.

\subsection{Magnetic field amplification and compression} 
\label{sec:dyn}

The turbulent magnetic field close to the shock can be enhanced by
several physical processes. Here we concentrate on the amplification
due to streaming instability induced by cosmic rays accelerated at the shock
and we show that this mechanism allows us to reproduce the strength of the
magnetic fields which are inferred by assuming that the thickness of the
X-ray filaments is due to severe synchrotron energy losses. On the other hand,
if the filaments are interpreted in a different way, 
very low efficiencies of proton acceleration are required in order to avoid 
effective magnetic field amplification.

The streaming of cosmic rays upstream of a shock excites both resonant
(\cite{skilling}) and non-resonant \cite[]{bell04} modes. The latter modes are
expected to dominate for fast shocks, efficiently accelerating cosmic rays,
most likely during the free expansion phase of the post-supernova evolution 
and at the very beginning of the Sedov phase \cite[]{kinetic}. At any 
later time, resonantly excited Alfv\'en waves should account for most of 
the field amplification in the shock vicinity. Although the predictions of
quasi-linear theory should be taken with caution when considering the regime 
of strongly non-linear field amplification, $\delta B/B\gg 1$, at
present no better treatment is possible. The strength of the magnetic 
field at the position $x$ upstream, $\delta B(x)$, in the absence of damping, 
can be estimated from the saturation condition, that, for modified
shocks, reads (\cite{long}):
\begin{equation} \label{eq:B_amp}
 p_w(x)= U(x)^{-3/2} \left[p_{w,0} +\frac{1-U(x)^2}{4\,M_{A,0}} 
            \right] \,,
\end{equation}
where $p_w(x)= \delta B(x)^2/(8\pi \rho_0 u_0^2)$ is the magnetic pressure 
normalized to the incoming momentum flux at upstream infinity. In 
Eq.~(\ref{eq:B_amp}), $U(x)=u(x)/u(_0)$ and $M_{A,0}=u_0/v_A$ is the
Alfv\'enic Mach number at upstream infinity where only the background
magnetic field, $B_0$, assumed to be parallel to the shock normal, is
present. $p_{w,0}$ accounts for the magnetic pressure due to a pre-existing
magnetic turbulence. This term is usually set to zero in the following 
calculations, unless otherwise specified.

Eq.~(\ref{eq:B_amp}) correctly describes into account the effect of compression 
in the shock precursor through the term $U(x)^{-3/2}$.
For the upstream temperature that we adopt in RX J1713.7-3946 (see below),
damping in the upstream region is expected to be negligible.

This expression has been derived for Alfv\'en waves, therefore the perturbation
is perpendicular to $B_0$, namely in the plane of the shock. The
magnetic field downstream of the subshock is further enhanced by compression:
\begin{equation} \label{eq:B_2}
 B_2= R_{sub} \, B_1,
\end{equation} 
where $B_1$ is the magnetic field immediately upstream of the subshock. 

The combined action of cosmic ray induced magnetic field amplification and
field compression at the subshock results in a situation in which the magnetic
pressure may easily exceed the pressure of the background plasma. As shown by
\cite[]{cap08}, this leads to a reduction of the compressibility of the plasma,
so that the value of $R_{tot}$ drops compared with the cases in which this
dynamical reaction of the magnetic field is ignored. As stressed by
\cite{cap08} this effect alleviates the need for invoking poorly known
processes such as the so-called turbulent heating. 

\subsection{Radiation produced by accelerated particles}
 \label{sec:photon}

In this section we briefly describe the calculations of the fluxes of
nonthermal radiation produced by electrons and protons in the SNR. 
We concentrate on the following channels: 1) gamma ray production by generation
and decay of neutral pions in inelastic $pp$ collisions; 2) synchrotron
emission of electrons; 3) ICS of electrons on the CMB and
infrared/optical photons in the remnant.

The flux and spectrum of gamma rays from $\pi^0$ decay are calculated
following the analytical approximations discussed by \cite{kel06}. 

The cutoff in the gamma ray spectrum resulting from $\pi^0$ decay can be
written as 
\begin{equation} \label{eq:nu_pi0}
 E_{\gamma,cut}^{\pi}\approx 0.15\, E_{\rm p,max}  \,,
\label{eq:cutgamma}
\end{equation}
where $E_{\rm p,max}$ is the maximum energy of accelerated protons.
Qualitatively, the spectrum of gamma rays at energy $E\ll E_{\gamma,cut}^{\pi}$
reproduces the shape of the parent proton spectrum. The gamma ray spectrum is
therefore expected to show some level of concavity, typical of acceleration at
cosmic ray modified shocks. At energies lower than the pion mass, the gamma ray
spectrum is found to rapidly drop. 

Synchrotron emission is produced by accelerated electrons in the magnetic field
amplified by streaming cosmic ray protons. The calculation of the synchrotron
emission is carried out by using the exact synchrotron kernel \cite[]{rybicki}.
The synchrotron spectrum is cut off at a photon energy
\begin{equation}
E_{syn,cut} \approx 3\times 10^{-3} B_{100}
\left(\frac{E_{e,max}}{\rm TeV}\right)^2 {\rm keV},
\label{eq:cutsyn}
\end{equation}
where $B_{100}$ is the local magnetic field in units of $100\mu G$. It is easy
to check that for $E_{e,max}\sim 10$ TeV, the synchrotron emission is cut off
in the keV energy range. It is also worth keeping in mind that in the
straightforward case of Bohm diffusion at a linear shock, the maximum energy
of electrons scales as $B^{-1/2}$, therefore $E_{syn,cut}$ is independent of
the magnetic field. In our calculations, the non-linear effects induce a weak
dependence on magnetic field, as discussed in \S~\ref{sec:results}.

In this paper we compute the synchrotron radiation spectrum limited to 
X-ray band. The radio emission, being produced by lower energy
electrons, with much longer lifetimes, is affected by the temporal evolution
of the supernova remnant, which strongly depends on the environment in which 
the blast wave propagates.

The X-ray emission is confined to a narrow region behind the shock, whose
thickness is determined, depending on the radiating electron energy, by either
the diffusion or advection length covered by electrons during their loss
time. As we show at the end of next section the contribution to the X-ray
radiation from the upstream region is negligible, due to the much lower
magnetic field (see Fig.~\ref{fig:f5} and the corresponding discussion).
The thickness of the radiating rim can be estimated by solving the
diffusion-convection equation in the downstream plasma: $u_2(\partial
f_e/\partial x)= D \partial^2 f_e/\partial x^2-f_e/\tau_{\rm syn}$, where
$B_2$ has been assumed to be spatially constant (i.e. we are neglecting any
magnetic damping mechanisms). The solution has the form $f(p,x)\propto \exp
(-x/\Delta R_{\rm rim}(p))$ where $\Delta R_{\rm rim} (p)$ is the spatial
scale of the emitting region for electrons with momentum $p$ and is given
by \cite[]{ber04}:
\begin{equation} \label{eq:rim}
 \Delta R_{\rm rim} = \frac{2 D_2/u_2}{\sqrt{1+  
  4 D_2 /u_2^2 \tau_{\rm syn}}-1}
 \,.
\end{equation}
The observed thickness of the rim is obtained by correcting the previous
expression for projection effects. In the ideal case of a spherical shock, an
exponential profile translates into a projected thickness $\Delta R_{\rm
obs}\simeq 4.6\Delta R_{\rm rim}$ \cite[]{par06}, where $\Delta R_{\rm
obs}$ is defined as the size of the region over which the observed brightness
drops to half of the maximum.

Finally, electrons contribute to gamma ray emission through their ICS of 
ambient photons. The main contribution to gamma ray emission comes from
upscattered dust-processed infrared photons, having a blackbody spectrum with
temperature $\sim 20$ K \cite[]{sch98}. Following the instance of the IR+Opt
photon background in the interstellar medium (ISM) we assume that the ratio of
the optical to infrared energy densities remains $\sim 20$, while the energy
density of IR light, $W_{IR}$, is left as a free parameter (in the ISM,
$W_{IR}\approx 0.05\,\rm eV cm^{-3}$).

The flux and spectrum of ICS photons is calculated by using the exact kernel
for ICS, with the full Klein-Nishina (KN) cross section \cite[]{rybicki}. The
spectrum is cut off at an energy
\begin{equation}
E_{ICS,cut}\approx 5 \left(\frac{E_{e,max}}{\rm TeV}\right)^2 
\left(\frac{\epsilon_{ph}}{\rm eV}\right)\, \rm TeV\,\,\,\,\,,
\label{eq:cutics}
\end{equation}
if $E_{e,max}\leq 260 (\epsilon_{ph}/\rm eV)^{-1}$ GeV and
$E_{ICS,cut}=E_{e,\rm max}$ otherwise. The boundary between the two regimes
corresponds to the setting in of the KN regime. For the photon background
we adopted here ($\epsilon_{ph} \approx 2\times 10^{-3}$ eV), the threshold
is at energy $\sim 150$ TeV, therefore for the case of RX J1713.7-3946, the
cutoff region falls well within the Thompson regime and Eq.~(\ref{eq:cutics}) 
applies.

\section{Results: RX J1713.7-3946} 
 \label{sec:results}

In this section we describe the results of our calculations for the case of 
SNR RX J1713.7-3946. A major difficulty in carrying out
detailed calculations on this remnant is that neither its distance nor its 
age are well established. \cite{koy97} infer a distance of $\sim
1$ kpc, based on ASCA X-ray observations, while \cite{sla99} estimate a
distance of $\sim 6$ kpc. Here we adopt the value of 1 kpc, which
appears to be supported by some more recent work \cite[]{fuk03,mor05}
based on the spatial correlation between the remnant and a molecular cloud
located at that distance. This estimate was also independently confirmed 
by \cite{cas04} based on the comparison between the interstellar
absorption as 
measured using X-ray and radio observations. Since the observed angular size 
of the remnant is $\sim 60'$, the remnant radius can be estimated to be 
$R_{\rm SNR}\simeq 10$ pc, which would lead to estimate an age of 
$\sim 1600$ yr, consistent with the historical Chinese record of a SN exploded 
in AD 393 \cite[]{wan97}.  

Several authors, based on different arguments, have reached the conclusion 
that RX J1713.7-3946 is a core collapse SN of type II/Ib, expanding into a 
hot and very diluted bubble created by the wind of a massive progenitor star 
with initial mass between $15<M<20\, M_\odot$ \cite[]{ber06,sla99,fuk03,cas04}. 
These authors also conclude that the bubble must have a radius of 
$15\lesssim R_b \lesssim 20$ pc, hence the shock front of the remnant is most
likely still propagating into the low density shocked wind, except perhaps
for the SW region where the shock seems to impact the cold and dense shell of
unshocked material. This scenario leads to a temperature for the bubble 
typically $\gtrsim 10^7$ K \cite[]{che89}. In order to account for the 
inhomogeneity of the medium, we adopt here a somewhat lower value of 
$T_0= 10^6$ K. We checked however that our results are very weakly
dependent on the value of the gas temperature at upstream infinity, as long as
it is larger than $\sim 10^4$ K.

Other uncertainties that affect the results of our calculations concern 
the density of the background medium, $n_0$, and the strength of the 
background magnetic field (before amplification takes place). Here we 
adopt $n_0=0.12\, \rm cm^{-3}$, which is somewhat dictated 
by the fitting to the HESS data.
Somewhat larger values of $n_0$ could be still compatible with HESS
observations if a fudge factor is introduced to account for the fact that
the gamma ray emission is not homogeneous over the SN shell. Several
attempts at measuring the ambient gas density have been made. An upper
limit of $\sim 2.6\, \rm cm^{-3}$ comes from the molecular hydrogen column 
density in parts of the SNR without detectable CO emission that can be 
derived based on NANTEN \cite[]{fuk03} measurements (see also \cite{aha06} 
for a discussion). 

The other existing constraint was inferred by \cite{cas04} based on the 
lack of thermal X-ray emission in XMM data. The limit reads $n_0<0.02\,\rm
cm^{-3}$. However, it is worth keeping in mind that this limit was 
derived by assuming that electrons and protons are in thermal equilibrium
downstream, but the same authors suggest that this is likely not the case. They
also investigate the non-equilibrium case but apparently they used an equal
temperature of electrons and protons in this second case as well.

An upper limit of $4500$ Km/s to the velocity of the ejecta was inferred by
\cite{uch07} based on a direct measurement of the transient X-ray emitting
region. On the other hand a value $u_0< 1500$ Km/s would not allow for
acceleration of electrons up to high enough energy to explain the observed 
X-ray emission. Here we adopt a value of $u_0=4300$ Km/s necessary to fit
the cutoff in the X-ray spectrum as detected by Suzaku. A lower value of
about$\sim 3000$ Km/s would be needed to fit the ASCA data, but we do not
discuss this case any further here, given the superior quality of the
Suzaku data.

\begin{figure}
\begin{center}
\includegraphics[angle=0,scale=1]{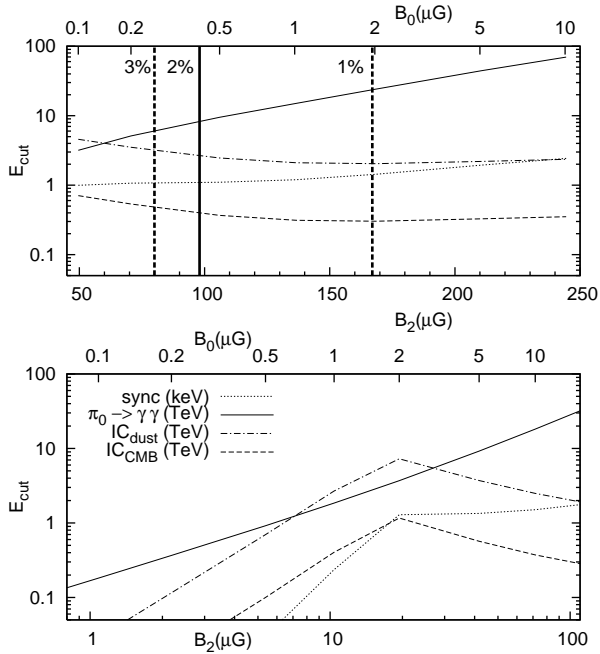}
\caption{Cutoff energies for photons produced by 
synchrotron emission (\textit{dotted} line, in keV), pion decay 
(\textit{solid}) and IC scattering (\textit{dot-dashed} for scattering on
IR radiation, \textit{dashed} for CMB), as a function of the downstream
magnetic field $B_2$. Also shown on the upper $x$-axis of both panels is 
the corresponding value of the background magnetic field $B_0$ that was 
used for the calculation. Notice that while the $B_2$-scale is logarithmic, the 
$B_0$-scale is arbitrary and built to show the correspondence 
between $B_0$ and $B_2$.
The \textit{upper panel} shows the case where $B_0$ is resonantly amplified
by the streaming instability. Here the thick vertical lines indicate the
solutions having the X-ray rim thickness equal to $1\%$, $2\%$ and $3\%$ of
the SNR radius respectively. In the \textit{lower panel} we assume no magnetic
field amplification, but simple compression of $B_0$ in the precursor and at 
the shock. The results in both panels are obtained with $\xi=3.7$
and $n_0= 0.12\, \rm cm^{-3}$, and with the maximum proton momentum determined 
by the condition $t_{acc}(p_{max})= 1600$ yr.}
\label{fig:f1}
\end{center}
\end{figure}

The model parameters are further constrained by the fact that the magnetic 
field in the downstream plasma, $B_2$, is an output of our calculations, for 
given $B_0$, as it is the result of cosmic ray induced magnetic field 
amplification and further enhancement by compression in the precursor and at 
the subshock. 
The correct solution of the problem is found by not only requiring that the
fluxes of observed radiations are fit, but also requiring that $B_2$
corresponds to the value inferred from the measured thickness of the
X-ray rim. For SNR RX J1713.7-3946, we use the data presented in \cite{laz04}
where the authors use a CHANDRA observation at 5 keV to infer the rim
thickness in the northwestern region of the SNR. The procedure leads to $\Delta
R_{\rm obs} \simeq 0.02 R_{\rm SNR}$ (see their Fig.~9). Clearly this thickness
does not lead to an estimate of $B_2$ if due to damping of the magnetic
field (as proposed by \cite{pohl05}) rather than to particles' energy losses.
Below, we first assume that the downstream magnetic field can indeed be 
derived from the extent of the X-ray filaments and then we investigate the 
consequences of relaxing this constraint. 

A first indication of the origin of the observed radiations can be obtained by
looking at the cutoff frequencies in the different bands. In the upper panel of
Fig.~\ref{fig:f1} we show the cutoff energies for synchrotron emission
(dotted line, in units of keV), $\pi^0$ decay (solid line, in TeV) and
ICS (dash-dotted line for ICS on the infrared light and dashed line for ICS 
on the CMB) as given by Eqs.~(\ref{eq:cutgamma}), (\ref{eq:cutsyn}) and
(\ref{eq:cutics}). The plot is obtained for fixed $\xi=3.7$ and varying $B_0$
between $0.1$ and $10\,\mu{\rm G}$ (upper $x$-axis), while the lower $x$-axis shows
the corresponding value of $B_2$. The vertical thick solid line shows the
solution that provides the central value of the measured rim thickness
(corresponding to $B_2= 100\,\mu$G) while the dashed vertical lines bound the
allowed region corresponding to $1\%<\Delta R_{\rm obs}/R_{\rm SNR}<3\%$. 

For the range of values of $B_0$ considered here, it is apparent that 
ICS cannot provide a high enough cutoff energy so as to explain HESS data. 
The latter require $E_{\gamma,cut}\gtrsim 10$ TeV, while the highest value 
of the ICS cutoff energy obtained for scattering on the IR light in the
presence of an extremely low background  magnetic field ($B_0 \sim 0.1 \mu
G$) is $\sim 4$ TeV. The corresponding cutoff energies for ICS off the
CMB photons are lower by  roughly the ratio of temperatures: $\sim
T_{dust}/T_{CMB}\sim 7$. 
The general trend of the ICS cutoff energy is to decrease as a 
function of $B_0$ (and $B_2$) because larger values of the magnetic field 
lead to stronger synchrotron losses and thereby lower maximum energies of 
the electrons. There are however two subtleties. First, taking into 
account nonlinear effects in the computation of the acceleration time reduces 
somewhat the dependence of $E_{\gamma,cut}$ on the magnetic field, with respect 
to the expectations of the linear shock analysis. Moreover, for values of $B_2$
lower than $\sim 20\ \mu$G, the maximum electron energy is determined by the
finite age of the remnant rather than by synchrotron losses. Hence, for 
values of the magnetic field below this threshold, the maximum electron 
energy increases with increasing magnetic field (because of the more effective
diffusion) and reaches a maximum for $B_2 \sim 20\ \mu$G. This corresponds 
to the highest possible value of $E_{\gamma,cut}$.

It is important to stress that the conclusion that the cutoff frequency for
ICS falls short of explaining HESS data by about one order of magnitude is
independent of arguments related to the normalization of the flux, which we
comment upon below.

The cutoff energy for gamma rays produced by $\pi^0$ decay is $>10$ TeV
provided $B_0>(0.3-1)\,\mu$G, which interestingly enough is also the region
where $B_2$ is consistent with the thickness of the X-ray rims, if these are
interpreted as the result of severe synchrotron losses downstream.

The cutoff energy of synchrotron emitted photons, which, assuming
Bohm diffusion, in linear theory would not depend on the value of the 
magnetic field, now acquires a weak dependence on $B_0$, and
varies between $\sim$ 1 and 2 keV for the considered range of magnetic 
field values, granted that $B_2 \gtrsim 20\,\mu$G. Things change
drastically  when $B_2$ is lower than this threshold and the maximum
electron energy is determined by the competition between acceleration time and
age of the system, rather than synchrotron losses (lower panel of
Fig.~\ref{fig:f1}).

The adopted value $\xi= 3.7$ corresponds to a fraction of injected
particles $\eta_{\rm inj}= 1.2\cdot 10^{-4}$. The resulting acceleration
efficiency, $\epsilon$, defined as the percentage of the energy flux crossing
the shock that ends up in relativistic particles, varies from $\sim 68\%$
(for $B_0=0.1\ \mu$G) to $\sim 36\%$ (for $B_0=10\ \mu$G).

It is very important to notice that the solutions corresponding to the curves
in the upper panel of Fig.~\ref{fig:f1} are characterized by moderately low
values of the compression factor in the precursor ($R_{tot}\lesssim 12$),
despite the high acceleration efficiencies. This result is purely the
consequence of the dynamical reaction of the amplified magnetic field on
the shock. No turbulent heating has been introduced in these calculations.
One may wonder how the results would change if no cosmic ray induced
magnetic field amplification were present, so that the magnetic field in the
precursor and at the shock would only be produced by compression of a turbulent
field at upstream infinity, with $\delta B\sim B_0$. In this case the
turbulent field (assumed to be perpendicular to $B_0$) is compressed
according to $B_\perp(x)= B_0 (u_0/u(x))$. In the bottom panel in 
Fig.~\ref{fig:f1} we show the cutoff energies obtained in this
situation (we still assume $\xi=3.7$). For $0.1<B_0<10$ $\mu$G, the total
compression factor ranges between 14 and 7, and $B_2$ reaches up to $\sim
100\,\mu$G. Remarkably, the highest value of the maximum energy of gamma rays
produced by ICS (off the IR photons) is $\sim 7$ TeV and is reached for
$B_0\approx 2\mu G$ ($B_2\approx 20 \mu G$). Lower and higher values of $B_0$
lead to smaller cutoff energy for the ICS emission. 

\begin{figure}
\begin{center}
\includegraphics[angle=0,scale=1]{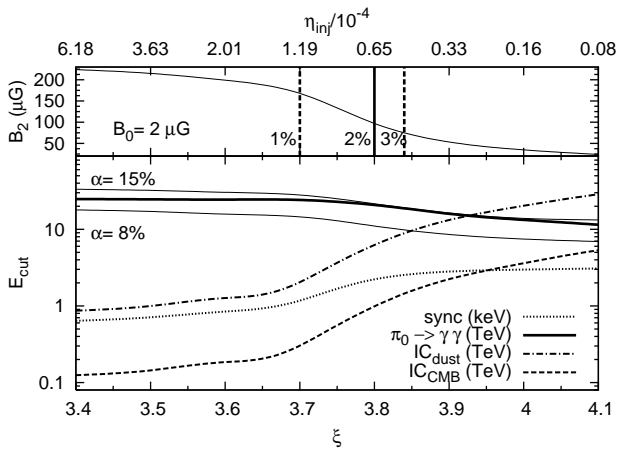}
\caption{The thick lines in the \textit{lower panel} show the cutoff
frequencies for the radiative processes (just as in Fig.~\ref{fig:f1}) 
now computed as a function of the injection efficiency $\xi$, for fixed
$B_0=2\ \mu$G. The thin solid lines show the cutoff frequency for $\pi^0$ 
decay when the maximum proton energy is computed according to the definition
on the right of Eq.~(\ref{eq:pmax}) with $\alpha= 0.08$ and $0.15$.
The downstream magnetic field, resulting from the amplification and
further compression of $B_0$, is shown in the \textit{upper panel}, where
the vertical lines have the same meaning as in Fig.~\ref{fig:f1}. The upper 
$x$-axis of this panel displays the fraction of injected particles in units 
of $10^{-4}$.}
\label{fig:f2}
\end{center}
\end{figure}

The results of the calculations are very sensitive to the recipe adopted 
for the injection of cosmic rays from the thermal pool. As discussed in 
\S~\ref{sec:max_energy}, we model the injection with the  {\it thermal
leakage-like} model introduced by \cite{bgv05}. In Fig.~\ref{fig:f2} we
illustrate the dependence of $B_2$ (upper panel) and of the cutoff energies
(lower panel) on the injection parameter $\xi$, varying between 3.4 and 4.1
(corresponding to $6\times 10^{-4} > \eta_{\rm inj} > 8\times 10^{-6}$).
The upstream magnetic field is $B_0=2\ \mu$G. The curves are labelled as in
Fig.~\ref{fig:f1}, with the addition of two thin solid lines representing
the cutoff energy for $\pi^0$ decay when the maximum energy of the parent
protons is calculated with $\alpha=8\%$ and $\alpha=15\%$ in
Eq.~(\ref{eq:pmax}).  The vertical thick solid line in the upper panel
shows the solution providing the central value of the measured rim
thickness (corresponding to $B_2= 100\,\mu$G) while the dashed vertical
lines bound the allowed region corresponding to $1\%<\Delta R_{\rm
obs}/R_{\rm SNR}<3\%$.

The cutoff energy for gamma rays due to ICS is again lower than $\sim 10$ TeV,
with the possible exception of ICS on IR light for the highest values of
$\xi$. For instance, for $\xi\sim 3.9$ the cutoff energy for ICS is $\sim
12$ TeV, but in this case the cutoff frequency for pion decay is at the
same level. In this limit case the issue becomes the absolute
normalization. In any case such situation also corresponds to $B_2\sim
50\mu$G, quite lower than inferred from X-ray measurements (compare with
the region between the two vertical dashed lines in the upper plot). 

\begin{figure}
\begin{center}
\includegraphics[angle=0,scale=1]{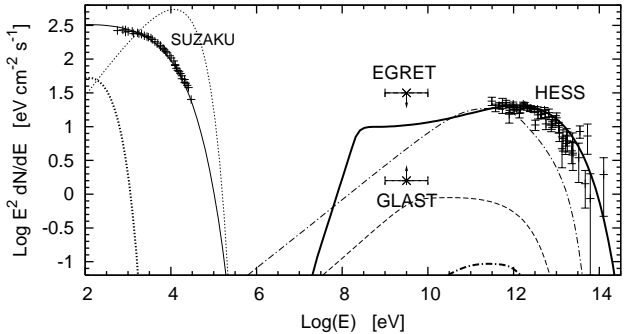}
\caption{Spatially integrated spectral energy distribution of RX
J1713.7-3946 in the hadronic scenario, for $n_0= 0.12\,\rm cm^{-3}$, 
$u_0=4300$ km/s, $B_0=2.6\, \mu G$, $\xi= 3.8$. The following components
are plotted: synchrotron (dotted line) thermal emission (dotted line),
Compton scattering with CMB (dashed line) and with Opt+IR background
(dot-dashed line). The contribution from pion decay is shown as a thick
solid line and corresponds to $p_{p,\rm max}= 1.26\times 10^5$. 
HESS data taken from 2003 to 2005 are plotted together with Suzaku data in
the X-ray band. Also EGRET upper limit and GLAST sensitivity are shown.}
\label{fig:f3}
\end{center}
\end{figure}

The above discussion leads to conclude that there is some degeneracy
in the choice of the values of the parameters of this problem. If we
try to fit the spectral energy distribution in the X-ray and gamma ray bands
and the thickness of the X-ray filaments simultaneously, we find that the best
fit is obtained for $1<B_0<3\,\mu$G and $3.7<\xi<3.85$. The resulting
acceleration efficiency varies in the range $20\%< \epsilon <60\%$. In 
Fig.~\ref{fig:f3} we show a possible solution with $B_0=2.6 \mu$G and
$\xi=3.8$. 
The contribution of $\pi^0$ decay to the gamma ray flux is plotted as a thick
solid line. The  solid curve in the lower energy part represents the
synchrotron emission of electrons, superimposed on the recently published
Suzaku data \cite[]{tan08}. The flux of ICS photons when the target is
made of CMB photons is plotted as a dashed line. The ICS on the IR+Opt
photons, assumed to have a density equal to the average value in  the
Galaxy ($1\,{\rm eV\,cm^{-3}}$) corresponds to the lower thick  dash-dotted
line. For illustration purposes, we also plot the contribution of ICS on a
IR+Opt background with energy density 200 times larger (higher dash-dotted
line). It is worth noticing that while synchotron emission is always the 
dominant loss mechanism for electrons, the rate of energy losses due to
inverse Compton scattering mainly comes from interaction with the optical 
photons. On the contrary, the leptonic gamma ray flux in the HESS band is 
mostly contributed by upscattering of the IR photons.

Fig.~\ref{fig:f3} shows a few important points: 1) in terms of shape,
normalization and position of the cutoff the contribution of $\pi^0$
decay provides a satisfactory fit to HESS observations; 2) the
ICS on the CMB photons is characterized by too low a normalization and 
too low cutoff energy; 3) for ICS on the IR+Opt light, the normalization 
can be raised to the observed flux level only if a very large photon flux 
is assumed; 4) even in this case the position of the cutoff in the gamma 
ray spectrum turns out to be below the observed one; 5) while being below 
the sensitivity limit of EGRET, the predicted fluxes should be observable 
by GLAST.

As the HESS observations show \cite[]{aha07}, the gamma ray spectrum in the
TeV region can be fit by a cutoff with shape $\exp[-(E/E_0)^\beta]$ with
$\beta$ close to 0.5. As shown here (see also \cite{kel06}), this result
is perfectly compatible with the spectrum of gamma rays from the decay of
neutral pions, if the cutoff in the spectrum of the parent protons is
exponential.

A few words are in order concerning the maximum momentum of  protons. The
value of $p_{p, \rm max}$ that provides the best fit to the data is
$1.26\times 10^5\, m_p c$, which is obtained when the maximum proton energy
is calculated by using the condition of finite size of the accelerator with
$\alpha=0.08$ in Eq.~(\ref{eq:pmax}) (see also Fig. \ref{fig:f2}). The
condition on the age of the remnant would have resulted in a maximum proton
energy larger by a factor $\sim 2$.

A criticism to the scenario just discussed could be raised in that the flux 
of thermal X-rays plotted in Fig.~\ref{fig:f3} as a thin dotted line exceeds 
the Suzaku observations. This is however true only if protons and
electrons in the downstream plasma are in thermal equilibrium, while there
are reasons to believe that this is not the case (\cite{cas04}). 
If one assumes $T_e= 0.01\,T_p$ the absence of thermal X-ray emission
is naturally explained (thick dotted line in the same plot).

The Suzaku  data points are fit extremely well by our calculations: 
a fact which at least confirms that the shape of the electrons spectrum 
close to the cutoff is $\propto \exp[-(E/E_{e,max})^2]$ rather than a simple 
exponential. In other words, the maximum energy of electrons in this remnant 
is determined by energy losses,  rather than by the finite age of the system.  
Although not very stringent, this argument suggests that the magnetic field 
cannot be too low, or otherwise  ICS losses would have to be dominant, implying 
the presence in the remnant of  a density of IR photons much larger than in
the ISM.

\begin{figure}
\begin{center}
\includegraphics[angle=0,scale=1]{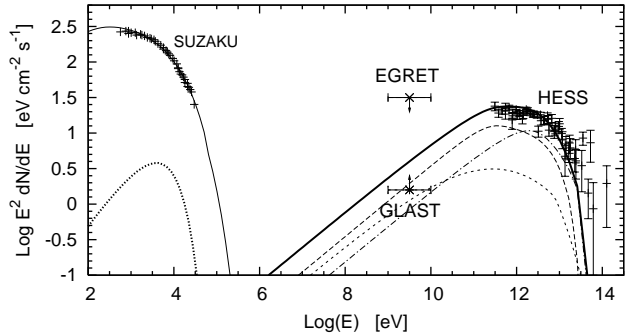}
\caption{Spatially integrated spectral energy distribution of RX
J1713.7-3946 in the leptonic scenario. The following components are
plotted: synchrotron (thin solid line) and thermal (thick dotted line) electron
emission, ICS component for CMB (dashed line), optical (dotted line) and IR
(dot-dashed line) photons, and the sum of the three (thick solid line).
The Opt+IR components are assumed to have energy density 24 times the mean ISM 
value. The other parameters are: $n_0= 0.01\,\rm cm^{-3}$,
$u_0=4300$ km/s, $B_0=1.5\, \mu G$ and $\xi= 4.1$.
The experimental data are the same as in Fig. \ref{fig:f3}}
\label{fig:f4}
\end{center}
\end{figure}

What if the narrow X-ray filaments are due to magnetic field damping or other
processes other than severe synchrotron losses? In this case there is no
constraint on the strength of the magnetic field at the shock, therefore one
may try to test the possibility of fitting the gamma ray data with the ICS emission
of high energy electrons. We carry out the previous calculations, including the
magnetic field amplification due to streaming instability but choosing now a
large value of the parameter $\xi$, so as to reduce the injection. This leads to 
two main effects: 1) lower efficiency of proton acceleration and 2) correspondingly
lower efficiency of magnetic field production. 

For the same values for the supernova as above (especially the distance and
velocity of the shock), and assuming $n_0=0.01\,\rm cm^{-3}$ and $\xi=4.1$,
we obtain the results illustrated in Fig.~\ref{fig:f4}. These parameters
correspond to a fraction of accelerated particles of order $\sim 7\times
10^{-6}$ and a cosmic ray energy conversion fraction of $\approx 2\%$. 
The curves are labelled as follows: the solid and dotted line on the
left side of the plot are the syncrotron and thermal electron emission 
respectively. The thin dashed, dotted and dot-dashed curves are the ICS on 
the CMB, optical and IR photons respectively, while the solid thick line is 
the sum of the three. The assumed energy density for the optical and IR 
radiation is 24 and $1.2\,\rm {eV/cm^{3}}$ respectively.

From Fig.~\ref{fig:f4} we conclude that a fit to HESS data on RX
J1713.7-3946 can be obtained within an ICS scenario at the price of: 1)
assuming an IR background in the SNR which exceeds the ISM value by a
factor $\sim 24$; 2) neglecting the highest energy data points of HESS.

The first issue has to do with the plateau-like shape of the gamma ray spectrum in
the lower energy part, which cannot be fit unless a second component of the
photon background (at higher frequencies) is assumed. This argument was also
made by \cite{tan08}.

The second point is due to the fact that despite the small efficiency for
proton acceleration, the magnetic field inferred for the downstream plasma is
still $\sim 20\mu G$, therefore the maximum energy of electrons is still not
large enough to have gamma ray emission above 10 TeV. 
Moreover it is worth noting that, in spite of the high energy density
assumed for the optical background, i.e. 24 eV/cm$^{-3}$, the synchrotron
losses still dominates over the IC losses because the scattering occurs in
the KN limit. As a consequence the spectrum of electrons is cut off as
$\exp[-(E/E_{e,max})^2]$, reflecting in a gamma ray spectrum with a rather
sharp cutoff which does not extend to sufficiently high energies to reach HESS
highest energy data. This is a point which is often overlooked in papers 
investigating the ICS interpretation of the HESS data.

\begin{figure}
\begin{center}
\includegraphics[angle=0,scale=1]{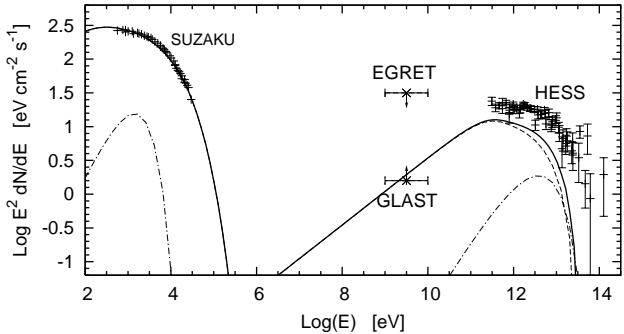}
\caption{Integrated spectral energy distribution for the leptonic scenario
as in Fig. \ref{fig:f4}. Here only the synchrotron emission and the ICS
for CMB are shown, separated in the contributions from the upstream
(dot-dashed line) and from the downstream region (dashed line). The solid
line is the total contribution.}
\label{fig:f5}
\end{center}
\end{figure}

The fluxes shown in Figs. \ref{fig:f3} and \ref{fig:f4} are obtained by
integrating over the emission volume, comprising the upstream and
downstream regions. It is instructive to show the emission in the different
bands from the upstream and downstream regions separately, since it
illustrates some of the non linear effects discussed above. In order to do
that, we concentrate on the case of inefficient acceleration ($\xi=4.1$)
which leads to the leptonic scenario. We stress again that even in this
case the magnetic field is generated upstream by the streaming cosmic ray
protons and further compressed in the precursor and at the subshock.
Moreover, for simplicity we use only the CMB photons as a target for ICS
(this is the reason why the predicted fluxes do not fit HESS data points).
The fluxes from downstream (dashed lines) and upstream (dot-dashed lines)
are illustrated in Fig. \ref{fig:f5}. The solid lines represent the total
emission, as the sum of the upstream and downstream contributions.  
For both synchrotron emission and ICS the main contribution comes from
the downstream region, at almost all photon energies. For synchrotron
emission, the relative ratio of the integrated emission from upstream
and downstream is larger than that calculated in the test particle
case (see for instance \cite{zir07}) because particles upstream feel
a smaller mean magnetic field in the non linear case, hence the
upstream synchrotron flux is smaller than in the test particle case. 
The contribution of electrons to ICS from the upstream region is also
smaller than the corresponding downstream contribution, although in
the cutoff region the former provides an appreciable fraction of the
total gamma ray flux.

\begin{figure}
\begin{center}
\includegraphics[angle=0,scale=1]{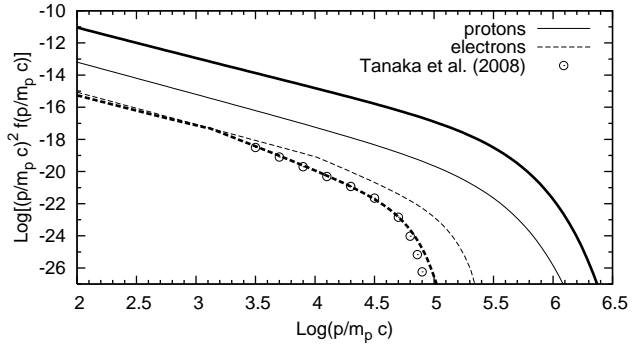}
\caption{Spatially integrated spectra for accelerated protons (solid lines)
and electrons (dashed lines) as functions of momentum. Thick and
thin lines refer to the hadron dominated scenario (see Fig.~\ref{fig:f3})
and lepton dominated scenario (see Fig. \ref{fig:f4}) respectively.
Circles represent the electron distribution as obtained by Tanaka et al.
(2008).}
\label{fig:f6}
\end{center}
\end{figure}

We conclude the review of the results of our calculations with
a discussion of the integrated spectrum of accelerated protons and
electrons. These are shown (multiplied by $p^2$) in Fig.~\ref{fig:f6}
for both the hadronic (thick lines) and the leptonic scenario (thin
lines). The break in the electron spectra is due to the onset of
synchrotron losses in reaching equilibrium. In both scenarios the
maximum energy of electrons is due to the balance between acceleration
and synchrotron losses, although in the leptonic case the situation is
borderline with the age of the system becoming comparable with the
time for synchrotron losses. 
The most important comment on the spectra of accelerated particles,
especially protons, is the concavity of the spectra which appears to
be barely visible. A standard non linear calculations with the shock
parameters adopted here would lead to strongly modified (concave) spectra.
The reduction of this effect is a crucial consequence of the dynamical 
reaction of the amplified magnetic field on the precursor, which leads
to a reduction of the spectral concavity (see discussion in \S
\ref{sec:dyn}). The cutoff in the spectrum of accelerated protons is
due to finite acceleration time. 
Fig.~\ref{fig:f6} also provides a visual estimate of the value of
$K_{ep}$ as it appears in Eq. \ref{eq:f_e(p)}, namely
$K_{ep}=6\times 10^{-5}$ for the hadronic scenario and $1.4\times
10^{-2}$ for the leptonic one.  
Finally, in Fig.~\ref{fig:f6} we also compare our electron energy
spectrum, as obtained in the hadronic scenario, with that derived by
\cite{tan08} (see their Fig.~12) directly from the Suzaku X-ray data
(circle points). The two spectra are in good agreement, with the
possible exception of the few data points at the highest energies.

\section{Conclusions and critical discussion}
 \label{sec:conc}

The detection of high energy gamma rays from SNRs has long been considered as
the most important tool to test the supernova paradigm for the origin of
galactic cosmic rays. We now have a couple of positive gamma ray detections
in the $1-100$ TeV energy range, but it remains unclear whether this emission is
of hadronic or leptonic origin. A hadronic origin would point to the
acceleration of cosmic rays up to energies close to the knee, which is the main
motivation for this work.

We investigated the problem in the context of the non-linear theory of particle
acceleration at SNR shocks, including the dynamical reaction of the accelerated
particles, the generation of magnetic fields as due to streaming instability
excited by cosmic rays, and the dynamical reaction of these self-generated
fields on the plasma. We specialized our calculations to the SNR RX
J1713.7-3946. Previous attempts to carry out similar calculations did not
include the magnetic field amplification (which was input {\it by hand} so
as to fit the data) and its dynamical reaction. The latter is crucial to reduce
the compression factors to the levels needed to explain the few available 
observations \cite[]{cap08}, while in previous work on the topic, this problem 
was faced by invoking turbulent heating (see for instance \cite[]{ber06}).

The non-linear calculations are applied to two cases with very different
acceleration efficiency. In the first scenario, the shock accelerates a 
fraction $\sim 10^{-4}$ of the incoming particle flux, which results in 
large acceleration efficiency and effective magnetic field amplification 
through resonant streaming instability. This scenario provides a benchmark 
for a hadronic interpretation of the HESS gamma ray observations of 
RX J1713.7-3946. In the second scenario a fraction $\sim 10^{-6}$ of the 
incoming particle flux through the shock gets accelerated. This serves as 
a model for a leptonic origin of the HESS data on RX J1713.7-3946.

The main discrimination between the two scenarios resides in the strength of
the magnetic field at the shock: efficient cosmic ray acceleration also leads
to the generation of magnetic field upstream of the shock which by compression
translates into a downstream field of $B_2 \sim 100\mu G$, compatible with the
value which has been inferred from the interpretation of the morphology of
the X-ray emission as due to synchrotron losses. The second scenario requires
that the magnetic field is very weakly amplified (if at all), so that $B_2\sim
15-25\mu G$. In this case the X-ray morphology is interpreted as a result
of damping of the magnetic field in the downstream plasma \cite[]{pohl05}. 

In the following we discuss these two scenarios in terms of their problems and
virtues, which also leads us to comment on the origin of the X-ray
filaments. 

Our calculations show that for the parameters typically adopted for RX
J1713.7-3946, and assuming $\xi=3.7$ for the injection, particle acceleration of
protons works efficiently. X-ray and gamma ray data can be fit very well within
this approach, and the strength of the downstream magnetic field turns out to
describe well the brightness profile of the X-ray emission without further
adjustments.
The cutoff in the gamma ray emission leads to conclude that a cutoff in the
parent proton spectrum is at $\sim 10^5$ GeV. We stress that this is the cutoff
in the proton spectrum which is being accelerated at the present time, while
the highest energy which can be achieved in a SNR is reached at the very
beginning of the Sedov phase. During the Sedov phase, particles which reached
higher energies at previous times escape the system from upstream (and indeed
these are the only particles which are allowed to leave the system from
upstream). The shape of the X-ray spectrum is taken from the recent Suzaku
observations of RX J1713.7-3946 \cite[]{tan08}, which lead to a cutoff
energy in the X-ray spectrum requiring a shock velocity of $4300$ km/s
within the framework of our model. 
For Bohm diffusion the energy of the X-ray cutoff is very weakly dependent on 
the magnetic field strength in the shock region (in fact in linear theory
it would have been independent of $B$). In this scenario the ICS
contribution from high energy accelerated electrons is negligible compared
with the contribution of $\pi^0$ decays, unless an unreasonably large
energy density of target IR light is adopted. 

The best fit is achieved by using a density of the upstream gas of $n_0=0.12 \,
\rm cm^{-3}$. The implications of this result are far reaching and deserve
some discussion: as shown in Fig. \ref{fig:f3} the emission expected from
the gas behind the shock as a result of thermal bremsstrahlung emission of
electrons exceeds the observed flux by a factor $\sim 6$. This conclusion could
be also derived based on simpler calculations, as done by \cite{kat08}.
However, as also argued by \cite{cas04}, the electrons and protons in this
SNR are not expected to be in thermal equilibrium because of the relatively
low density of the medium. This possibility is also discussed by
\cite{ghav07}, who find that for a remnant with velocity $3000$ km/s the
ratio of the electron to proton temperature is $\sim 0.02$ and scales as
$\sim V_s^{-2}$ (recall however that these conclusions were obtained in the
context of linear calculations, and could be even stronger if the
non-linear effects are taken into account). We showed that such a 
temperature ratio would lead to predict a thermal X-ray emission well below
the sensitivity of Suzaku and ASCA. On the other hand, although the age of
this remnant is too short to allow Coulomb scattering to equilibrate
electrons and protons downstream, one should check that Coulomb scattering
acting for $\sim 1600$ years does not heat electrons enough (about 1 keV)
to excite lines in iron atoms, which would then become easily visible on
top of the synchrotron emission (see for instance \cite{many} for a
discussion of this effect).

\cite{kat08} also criticized the hadronic scenario based on the lack of
an excess gamma ray emission in the direction in which, according to
\cite{cas04} the shock appears to have impacted a nearby molecular cloud,
where the target density for $pp$ collisions should be higher. As simple as this
inference may sound, it is not that straightforward: the gas in the molecular
cloud is expected to be mostly neutral, a fact that has at least two important
consequences. First, only the ionized component takes part in the acceleration
process (recall that in order to be scattered back towards the shock from
downstream, incoming particles should be ionized within a few times the shock
thickness). This implies that the efficiency of particle acceleration in the
proximity of the molecular cloud could be much lower than in less dense parts
of the surrounding medium. Moreover, in a mostly neutral medium, the Alfv\'enic
turbulence on which efficient particle acceleration relies can be effectively 
damped, thereby leading to a lower maximum energy of accelerated particles. 
These two instances show that having a dense target for $pp$ collisions does 
not necessarily imply that there should be a larger gamma ray flux from 
$\pi^0$ decay. 

As already stressed, the actual trademark of the hadronic scenario discussed
above is the prediction that the downstream magnetic field in RX J1713.7-3946
is $B_2\sim 100\mu G$, which would imply that the filaments observed in X-rays
are due to severe synchrotron losses of accelerated electrons. \cite{pohl05}
argued that a narrow region of X-ray emission could also be due to damping of
the magnetic field downstream of the shock. The argument was also adopted by
\cite{kat08} in order to suggest that HESS gamma rays must be of leptonic
origin. \cite{kat08} also quote the results of \cite{roth} on SN1006 to stress
that filamentary structures are observed also in the radio and not only in
X-rays, therefore the filaments cannot be due to synchrotron losses. It should
be noticed, however, that \cite{roth} use their data to reach the opposite
conclusion. In fact \cite{roth} plot the spatial profiles of the radio and
X-ray emission, which show that the radio emission profile has only a shallow
bump, when compared with the pronounced peak in the X-ray brightness. We should
recall that the thickness of the damping region, as correctly proposed by
\cite{pohl05}, is determined by using the convection equation for the waves,
with no generation terms and only a damping term. This trivially leads to a
solution which is an exponential in the spatial coordinate with a cutoff scale
which defines the damping scale. It is worth reminding that the exponential
shape is the very reason why one should expect a filament. In order to avoid
that an equally pronounced filamentary structure appears in radio,
\cite{pohl05} are forced to assume that the field has a residual large scale
component which is not affected by the damping: the radio emission then comes
from lower energy electrons propagating in this residual field. This solution
appears to us rather fine tuned, a limitation that should be kept in mind when
discussing the damping explanation of the X-ray morphology. This limitation is
probably even more serious in the case of RX J1713.7-3946, where the X-ray 
emitting region is comparatively broader and it is not very easy to explain 
it with the damping models proposed by \cite{pohl05} (see Eqs.~(7) and (12) 
of their paper), although there are many parameters in their formulae that 
can be appropriately tuned. 

In our opinion, a rather serious concern about a hadronic origin of
the HESS gamma ray emission, and indeed about this remnant being an
efficient cosmic ray accelerator, is the low value of $K_{ep}$ predicted by the
non-linear theory. Our calculations lead to $K_{ep}\sim 10^{-4}$, which is much
lower than the value inferred from direct measurement of the electron spectrum
in the $10$ GeV energy region, at the Earth. At first sight one should conclude
that if the gamma radiation detected by HESS is of hadronic origin, then
electrons should have a different origin, but a dedicated investigation of the
problem should be planned to address this issue, since the very problem of
electron escape from the remnant needs to be carefully addressed. One obvious
warning against easy conclusions in this respect is that the value of $K_{ep}$
measured at the Earth is not the same as inferred for a single SNR at a given
age. It is rather an average over the temporal evolution of a remnant
and a sum over the many SNRs that contribute a flux at any given time. In fact
$K_{ep}$ is in general a function of time for a SNR, and it is possible that low
energy electrons are more effectively produced at later times when proton
acceleration and magnetic field amplification become less efficient, and
probably there is no appreciable gamma ray emission due to pp scattering.

Now we discuss our attempt to fit the HESS data with a leptonic model.
We stress that the calculations carried out here are exactly the same as
described above, with the only difference that we assume a larger value for the
injection parameter $\xi$, chosen here to be $\xi=4.1$ (corresponding to a
fraction of accelerated particles $\sim 10^{-6}$). The generation of magnetic
field by streaming instability is also kept in the calculations, although the
predicted field strength is clearly much lower than in the previous case. 

As shown in Fig.~\ref{fig:f4}, a fit to the combined HESS and Suzaku data is
possible if the highest energy HESS data points are ignored. The cosmic 
ray acceleration efficiency corresponding to this solution is $\sim 2\%$. The 
magnetic field found as a result of streaming instability and further 
compression in the precursor and at the subshock is $\sim 20\mu G$ and the 
cutoff energy of the electrons is still determined by energy losses.
This is an important point, because the cutoff in the electron spectrum is
super-exponential (namely $\propto \exp[-(E/E_{cut})^2]$) only if the rate
of losses is $\propto E^2$. If the maximum energy of electrons is
determined by the finite age of the accelerator, then an exponential cutoff
should be expected. The two situations reflect in profoundly different
X-ray spectra in the cutoff region. As seen in Figs.~\ref{fig:f3} and
\ref{fig:f4}, the Suzaku data are well fit by an electron spectrum with a
super-exponential cutoff. 

The fit illustrated in Fig.~\ref{fig:f4} has two problems, typical of leptonic
models: 1) the nearly flat region in the HESS data cannot be well fit by using
CMB photons alone as the target for ICS, as also discussed by \cite{tan08}.
In order to solve this problem, a population of IR photons needs to be
assumed, with energy density $\sim 20$ times larger than in the ISM. We
consider this as a problematic aspect of the leptonic model, since this
population of photons should be introduced {\it ad hoc}, without a proper
independent physical motivation. 2) The highest energy data points in the
HESS data cannot be fit with ICS of electrons. These data points have been
curiously ignored in some of the previous literature on the topic, probably
because of their relatively large error bars. It is however not easy to
imagine how the highest energy data point, at $E\sim 100$ TeV may
disappear. Future observations of this remnant in the multi-TeV range will
tell us if this might be the case. 

Figs.~\ref{fig:f3} and \ref{fig:f4} clearly show that a major discrimination
power between the two scenarios can be achieved with GLAST. In the region
$1-100$ GeV, the two scenarios imply a flux which is different by about one
order of magnitude. A different conclusion was reached by \cite{yama}, due to
an oversimplified treatment of particle acceleration in the non-linear regime,
which ignored several effects (magnetic dynamical reaction, turbulent heating)
leading to much less concave spectra of accelerated protons.

\section*{Acknowledgments}
We are grateful to Damiano Caprioli and Gamil Cassam-Ch\"enai for their
precious comments. We are also grateful to the anonymous referee for
the insights that contributed to improve the paper.
This work was partially supported by INAF (under grant PRIN-2005), by
MIUR (under grant PRIN-2006), by ASI through contract ASI-INAF I/088/06/0
and (for PB) by the US DOE and by NASA grant NAG5-10842. Fermilab is
operated by Fermi Research Alliance, LLC under Contract No.
DE-AC02-07CH11359 with the United States DOE.

\end{document}